# Magnetic Proximity Effect in 2D Ferromagnetic CrBr$_3$/Graphene van der Waals Heterostructures

Chaolong Tang,[†] Zhaowei Zhang,[†] Shen Lai,[†] Qinghai Tan,[†] Wei-bo Gao*,[†,‡]

[†]Division of Physics and Applied Physics, School of Physical and Mathematical Sciences, Nanyang Technological University, Singapore 637371, Singapore

[‡]The Photonics Institute and Centre for Disruptive Photonic Technologies, Nanyang Technological University, 637371 Singapore, Singapore

## ABSTRACT

Two-dimensional (2D) van der Waals heterostructures serve as a promising platform to exploit various physical phenomena in a diverse range of novel spintronic device applications. The efficient spin injection is the prerequisite for these devices. The recent discovery of magnetic 2D materials leads to the possibility of fully 2D van der Waals spintronics devices by implementing spin injection through magnetic proximity effect (MPE). Here, we report the investigation of magnetic proximity effect in 2D CrBr$_3$/graphene van der Waals heterostructures, which is probed by Zeeman spin Hall effect through non-local measurements. Zeeman splitting field estimation demonstrates a significant magnetic proximity exchange field even in a low magnetic field. Furthermore, the observed anomalous longitudinal resistance changes at the Dirac point $R_{XX,D}$ with increasing magnetic field at $\nu = 0$ may attribute to the MPE induced new ground state phases. This MPE revealed in our CrBr$_3$/graphene van der Waals heterostructures therefore provides a solid physics basis and key functionality for next generation 2D spin logic and memory devices.

**INTRODUCTION**

The exploration of the spin degree of freedom holds promise for next-generation memory and logic devices[1,2]. These devices use spins to carry information instead of charges and therefore can lead to low-power, high-speed operations[3]. Graphene provides an interesting platform for spintronics with the unique advantages of a long spin lifetime and long spin transport distance[4,5]. For spintronic devices, spin initialization is a prerequisite. There are three main methods for spin injection in graphene[6]. First, conducting ferromagnetic electrodes directly on graphene can lead to efficient spin injection. However, it suffers from the drawback of naturally short-circuit the graphene layer and restrict the design of spin switches. Second, magnetic dopants can be introduced into graphene. It can lead to the degradation of its unique property of high carrier mobility. Therefore, a lot of efforts have been devoted to the third method, that is, the introduction of magnetism in graphene with the proximity effect in the interface of magnetic insulators and graphene. Pioneer works have been demonstrated with 3D bulk magnetic insulators like yttrium iron garnet, Europium (II) sulfide and bismuth ferrite[7-13].

Due to the short-range nature of the magnetic exchange coupling, a fully 2D van der Waals heterostructures is desired for downsizing the device and introducing magnetic proximity effect (MPE) at the same time. The recent profound discoveries of 2D ferromagnets[14-27] bring the possibility of 2D ferromagnetic van der Waals heterostructures[28-34]. It is urgent and essential to investigate the magnetic coupling between graphene and these 2D ferromagnetic materials for developing 2D spintronic devices. Here, we report the investigation of MPE in CrBr$_3$/graphene van der Waals heterostructures, which is probed by Zeeman spin Hall effect through non-local measurements. A significant MPE field can be introduced even in a low magnetic field. Furthermore, we observed anomalous longitudinal resistance changes at the

Dirac point $R_{XX,D}$ with increasing external magnetic field at $\nu = 0$. This may attribute to the MPE induced ground state phases transformation of graphene from the ferromagnetic state at the lower magnetic field and a canted antiferromagnetic state at a higher field in quantum Hall regime.

A typical CrBr$_3$/graphene van der Waals heterostructure with Hall bar structure for electrical transport measurement is fabricated as shown in Figure 1a. In order to reach the best performance and a substantial MPE in CrBr$_3$/graphene heterostructures, we optimize the fabrication process and conditions to achieve the desired heterostructures (see methods for details). The sample in the final stage is encapsulated and protected by PMMA, which keeps the device surface away from the moisture and air for cryotemperature tests. The atomic structure of layered CrBr$_3$ is shown in Figure 1b. The Cr$^{3+}$ ions are configured in a honeycomb network, and the green arrows represent the spin direction of Cr atoms, which are found to exhibit a strong ferromagnetic coupling[35-40]. The Raman spectra of Figure 1c indicates the monolayer graphene still preserves a high crystal quality when heterostructured with CrBr$_3$ layer. The optical image in Figure 1d displays the as-fabricated CrBr$_3$/graphene heterostructures, along with the H-shape Hall bar structure for conducting non-local measurements.

As shown in Figure 2a, with the Zeeman effect, the splitting of the Dirac cone in graphene generates electron- and hole-like carriers with opposite spins near the Dirac point. This property can induce Zeeman spin Hall effect, with the working principle shown in Figure 2b. First, a source-drain current $I$ is applied. In the presence of an external perpendicular magnetic field, spin-up electrons and spin-down holes will be driven into the same direction with the Lorentz force, and hence no net transverse charge current is produced in the sample. However, this will generate a non-zero net spin current (Figure 2b). With the configuration, a non-local voltage $V_{nl}$ can be measured in the remote regions. With a larger magnetic field, we expect to observe a larger $V_{nl}$. Therefore, we could probe the introduced MPE in CrBr3/graphene heterostructures via the Zeeman spin Hall effect (ZSHE) in graphene.

The observation of MPE in CrBr3/Graphene heterostructures through ZSHE is shown in Figure 2c,d. The acquired non-local resistances $R_{nl} = V_{nl}/I$ as a function of the back-gate bias $V_g$ under a series of external fields are plotted in Figure 2c. The $R_{nl}$ peak at the Dirac point is written as $R_{nl,D}$. Along with the external magnetic field increasing, the non-local resistance $R_{nl,D}$ rises sharply, suggesting a distinct Zeeman splitting energy enhancement. Moreover, a few other possible sources of producing non-local observation like Ohmic contribution were analyzed and excluded (Supporting Information S5). In order to confirm the origin of this Zeeman splitting energy enhancement resulting from the MPE, we further measured the temperature dependence of $R_{nl,D}$. Figure 2d indicates that $R_{nl,D}$ in CrBr3/graphene heterostructures experiences an abrupt slowdown as the temperature exceeds the transition temperature $T_C$ (~37 K) of CrBr3, which induces the CrBr3's phase change from ferromagnetic to paramagnetic state. In addition, relevant literature shows that graphene has very weak temperature dependence when coupled with other non-magnetic materials like graphene/ AlOx[11]. Undoubtedly, we can therefore conclude that the MPE is the dominating contribution to induce $R_{nl,D}$ in CrBr3/graphene heterostructures when the temperature is lower than the $T_C$ of CrBr3.

Using as described Zeeman spin Hall probing, magnetic field dependence measurement and related theoretical calculation of the magnetic field induced Zeeman splitting energy in CrBr3/graphene could provide quantitative information on MPE in CrBr3/graphene heterostructures. Figure 3a is the plot of normalized $R_{nl,D}$ (using $R_{nl,D}$ measured at 4 T as reference) with external magnetic field sweep. It can be observed that $R_{nl,D}$ reaches orders-of-magnitude increase with respect to the magnetic field from 0 to 4 T. Inset curve is the measured $R_{nl}$ as a function of $V_g$ under 4 T. The quantitative description of MPE in CrBr3/graphene heterostructures can be estimated by calculating its Zeeman splitting energy $E_Z$ and the total Zeeman field $B_Z$. We firstly adopt the definition of $R_{nl}$ as follows[41-44]

$$R_{nl} \equiv \frac{dV_{nl}}{dI} \propto \frac{1}{\rho_{xx}} \left(\frac{\partial \rho_{xy}}{\partial \mu} E_Z\right)^2 \tag{1}$$

where $\rho_{xx}$ and $\rho_{xy}$ is the longitudinal and Hall resistivity, respectively, both are strongly depending on the external out of plane magnetic field ($B_\perp$); $\mu$ is the chemical potential, $E_Z$ is the Zeeman splitting energy which is proportional to the total external field ($B$). Therefore, we have $R_{nl} = \beta(B_\perp)B^2$, $\beta$ represents the $\rho_{xx}$ and $\rho_{xy}$ manifested parameters. The $R_{nl}$ peak at the Dirac point $R_{nl,D}$ could be further written as $R_{nl} = R_0 + \beta(B_\perp)E_Z^2$, $R_0$ stands for the non-local resistance of non ZSHE signal at zero field. Moreover, the $E_Z = g\mu_B B_Z = g\mu_B(B_{MPE} + B(\mu_0 H))$ can be further written as (Supporting Information S1)

$$E_Z = E_{Z0} \cdot \frac{\sqrt{(1+\mu_*^2 B^2)B_0^2}}{\sqrt{(1+\mu_*^2 B_0^2)B^2}} \cdot \frac{\sqrt{R_{nl,D}(B)-R_0}}{\sqrt{R_{nl,D}(B_0)-R_0}} \tag{2}$$

where $\mu_*$ is the carrier mobility, $B_0$ is the zero magnetic field, $B_{MPE}$ represents the field induced by MPE in CrBr$_3$/graphene heterostructures. Since a few layers and bulk CrBr$_3$ show zero polarization or magnetization near 0 T[38], we could reasonably neglect the MPE field and approximate $E_{Z0} = g\mu_B B_0$ by using $B_0 = 0.1$ T as reference magnetic field for calculation convenience to estimate the $E_Z$ in low field. The calculated Zeeman splitting energy $E_Z$ and its corresponding Zeeman field $B_Z$ are plotted against the external magnetic field in Figure 3b, respectively. The inset provides field effect mobility[45,46] ($\mu_*$) versus carrier density of CrBr$_3$/graphene heterostructures, high carrier mobility ensures the excellent device performance of 2D magnetic coupling and further observation of pronounced quantum oscillations. Figure 3b shows a clear evidence of the substantial magnetic exchange field by presenting two different slopes of line segments as well as the magnetic field increasing. When the applied field is low and less than about 1 T, the line slope is steeper than the one with field larger than 1 T, which demonstrates that magnetic exchange field is the dominating contribution to the nonlocal resistance at low field. Giving $\mu_0 H = 4$ T, the estimate $B_Z$ is approximately twice of applied field, which seems weaker than other reported 2D or graphene/ferromagnetic material systems like $WSe_2/EuS$[47], Graphene/EuS[11] and Graphene/$BiFeO_3$[12], but this result is reasonable, because 2D $CrBr_3$ has a smaller interlayer exchange coupling energy $J = 1.56$ meV[48], and its magnetic moment per Cr ion is 3.25 $\mu_B$[49], comparing with EuS which has an exchange coupling energy $J \sim 10$ meV, and the magnetic moment per Eu ion is 7.9 $\mu_B$[50].

As shown in Figure 4, four-probe longitudinal resistivity measurement further reveals the MPE effect in the quantum Hall regime of $CrBr_3$/graphene heterostructures, which not only lifts the ground state degeneracy of graphene but also indicates the ground state phase transformation near $v = 0$ in the quantum Hall regime. Color map of Figure 4a shows typical Landau Fan of the longitudinal resistance $R_{XX}$ as a function of the gate bias $V_g$ and the applied magnetic field $B$[51]. The bright red region at Dirac point suggests that a gap is opened. We further extract longitudinal resistance measured under external field from 2 to 9 T as shown in Figure 4b. For clarity of labelling the Landau level filling factors, resistance curves are vertically shifted and are proportional to the applied magnetic field. The filling factor $v$ of the Landau level can be obtained by using $v = \pm 4(|n|+1/2)$ where $n = 0, 1, 2 \ldots$ is the Landau level index. As shown in Figure 4c, $v$ can be directly derived from the longitudinal resistivity versus gate bias curve at each local $R_{XX}$ minimum, which shows pronounced quantum oscillations. Inset figure manifests that $V_g$ is linearly proportional to each $v$ value, as expected. Since the injected carriers by gate voltage into graphene could be expressed with $4e^2B/h$, we could use capacitance model to infer the Dirac point position to be about 8.5 V by considering gate voltage at $v = \pm 2$, as the result shown in Figure 4d. Most importantly, the close-up of each local $R_{XX}$ maxima and $R_{XX}$ minima of resistance curves taken at 2 T, 3 T and 4 T in Figure 4e reveal an anomalous resistance dip and peak-splitting features near $v = 0$ that develop with the increasing magnetic field, suggesting the lifting of Landau level degeneracy under large $B_Z$ with the contribution from MPE between CrBr$_3$ and graphene. Furthermore, we observed the anomalous longitudinal resistance at the Dirac point $R_{XX,D}$ with increasing magnetic field (Figure S1), which results from the MPE induced $B_Z$ in CrBr$_3$/graphene heterostructures, and attributes to the ground state phases of graphene transition from a ferromagnetic state to a canted antiferromagnetic state at $v = 0$ (Supporting Information S2).

In summary, we report the investigation of MPE between monolayer graphene and 2D ferromagnet CrBr$_3$, which results in a considerable Zeeman splitting field in graphene. The MPE effect shown in our CrBr$_3$/graphene van der Waals heterostructures therefore not only reveal the origin and prediction of magnetic exchange field at the interface of 2D van der Waals heterostructures, but also guide the design of next-generation 2D spin logic and memory devices via effect control of local spin generation and modulation.

**Experimental Methods**

*Device Fabrication:* The natural graphite and CrBr$_3$ crystals were purchased from HQ graphene. First, a monolayer graphene sheet was exfoliated and transferred onto a 285 nm SiO$_2$/Si substrate by using conventional scotch tape mechanical exfoliation method, the thickness of monolayer graphene was identified by its optical contrast on SiO$_2$/Si substrate with an optical microscope and then confirmed by Raman spectroscopy. For electrical transport measurement of CrBr$_3$/graphene van der Waals heterostructures, subsequently, we use electron-beam lithography, metal deposition and lift-off process to deposit Cr/Au electrodes (10 nm/20 nm) on the graphene layer, followed by using electron-beam lithography again and oxygen plasma etching to form Hall bar structure of graphene. Both acetone and isopropanol washing followed by a vacuum annealing were adopted to remove the PMMA residues from Graphene after contact formation and Hall bar patterning, this step is important because any residue on the graphene surface will not allow a good interface between $CrBr_3$ and graphene. Note that during the heterostructures assembly and contact processing, $CrBr_3$ is not initially stacked on graphene layer because the electrodes fabricating process (especially the electron-beam exposure, plasma etching and subsequent lift-off process) would seriously damage the $CrBr_3$. As to obtain a few $CrBr_3$ layers, the $CrBr_3$ crystal was first exfoliated onto a polydimethylsiloxane (PDMS) stamp in an inert atmosphere glove box with less than 0.1 ppm water and oxygen. Similarly, very thin $CrBr_3$ layers of different thicknesses were identified by using the optical contrast under different color filters and dark-field imaging. After that, a 10 nm thick $CrBr_3$ layer was picked up by a PDMS stamp in the glove box and transferred into the graphene Hall bar channel. Finally, the protective and insulated thin poly(methyl methacrylate) (PMMA) membrane was coated onto the $CrBr_3$/graphene/substrate to achieve our desired $CrBr_3$/graphene van der Waals heterostructures. Here it is important to choose a $CrBr_3$ layer which fits the whole Hall bar channel to guarantee it also has good contact with graphene near the contact edges.

*Electrical Measurements:* Transport and magnetotransport measurements were performed in a Physical Property Measurement System (Model 6000) down to 2 K and up to 9 T. The nonlocal and four-terminal longitudinal electrical resistances were acquired using a Keithley 2182A nanovoltmeter and a Keithley 2636B source meter with excitation current of 0.5 μA for high accurate electrical measurements. Multiple devices were prepared and measured (Supporting Information S3,4).

ASSOCIATED CONTENT

**Supporting Information**

The Supporting Information is available free of charge on the ACS Publications website at DOI.

Additional information on Zeeman splitting energy and field calculation, ground states of graphene in CrBr$_3$/graphene heterostructures, other samples' measurements and possible sources of producing nonlocal resistance $R_{nl}$.

AUTHOR INFORMATION

**Corresponding Author**

**\*E-mail: wbgao@ntu.edu.sg.**

**Notes**

The authors declare no competing financial interest.

Author Note: During the preparation of this manuscript, we became aware of a recent report on first-principle calculations of graphene/CrBr$_3$ van der Waals heterostructure[52] and another report on Hanle spin precession measurements of Cr$_2$Ge$_2$Te$_6$/graphene heterostructure[53].

ACKNOWLEDGMENTS

Special thanks to Stephan Roche for his help and informative discussion. This work was supported by Singapore NRF fellowship grant (NRF-NRFF2015-03) and NRF QEP grant, Singapore Ministry of Education (MOE2016-T2-2-077, MOE2016-T2-1-163 and MOE2016-T3-1-006 (S)), A\*Star QTE programme.

**Figures**

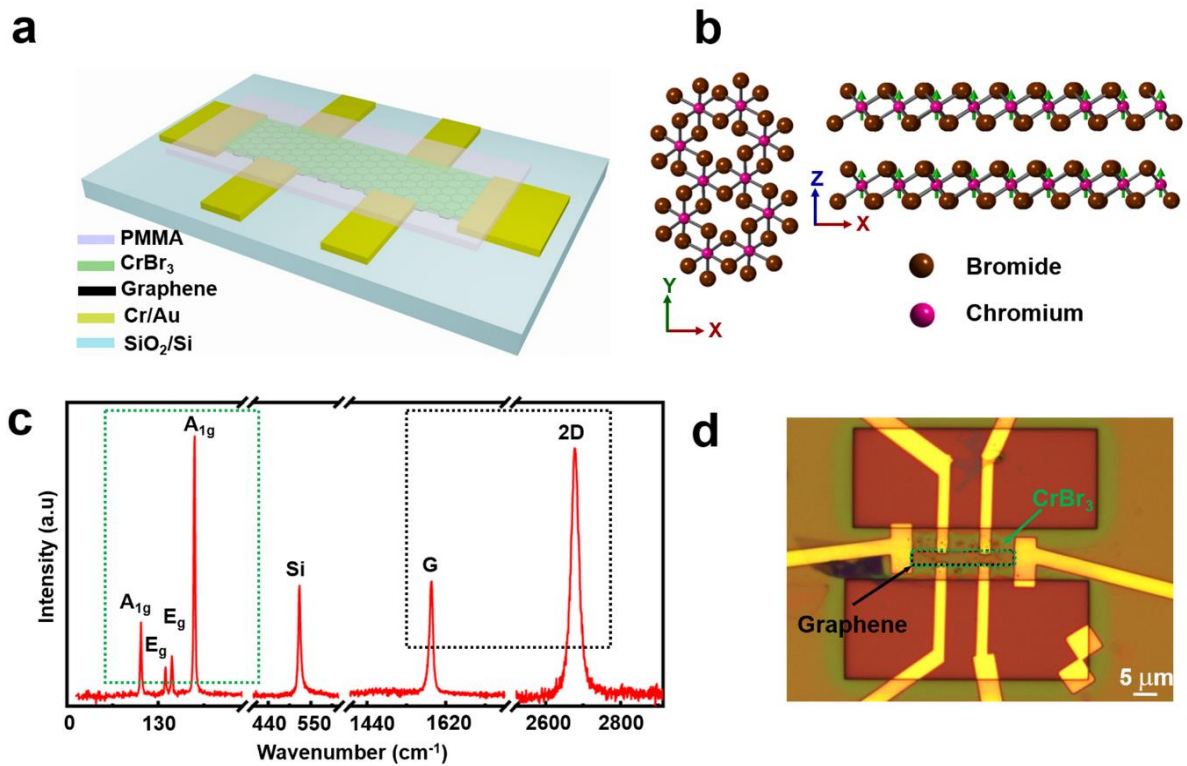

**Figure 1**. Fabrication and characterization of 2D ferromagnetic $CrBr_3$/graphene van der Waals heterostructures. (a) Scheme of $CrBr_3$/graphene heterostructures configuration. (b) Atomic structure of very thin $CrBr_3$ layers. Purple atoms: chromium; Brown atoms: bromide. Green arrow presents the spin direction of $Cr^{3+}$ atoms. (c) Raman spectrum of as-fabricated $CrBr_3$/graphene van der Waals heterostructures without top protective cover (the black dashed area is graphene; Green dashed area is $CrBr_3$). (d) Optical image of the as-prepared $CrBr_3$/graphene heterostructures. The scale bar is 5 μm.

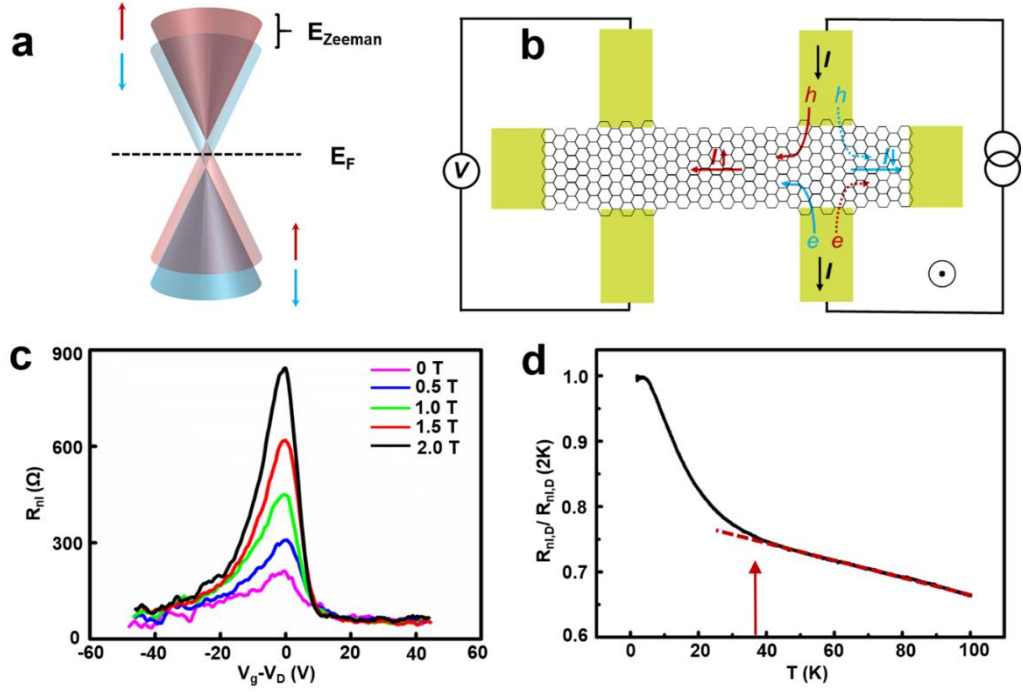

**Figure 2**. Zeeman spin Hall effect and non-local measurements in CrBr$_3$/graphene heterostructures. (a) Schematic diagram of Zeeman splitting of the Dirac cone in monolayer graphene (arrow represents the carrier spin direction). (b) Working principle of typical non-local measurements for probing Zeeman spin Hall effect. The driven current produces transverse spin up (red) and spin down (blue) currents under an external perpendicular magnetic field. Because the force has opposite signs for electrons and holes, the net charge current is zero, whereas the net spin current is nonzero. The resulting imbalance in the up/down spin distribution can reach remote regions and generate a voltage drop $V$. (c) The acquired non-local resistance $R_{nl}$ as a function of the back gate bias $V_g$ under a series of external field for this CrBr$_3$/graphene heterostructures at temperature T = 2 K. The $R_{nl}$ peak at Dirac point written as $R_{nl,D}$, the non-local resistance $R_{nl,D}$ in CrBr$_3$/graphene heterostructures rises sharply along with the external magnetic field increasing, suggesting a distinct Zeeman splitting energy enhancement. (d) Temperature dependence of $R_{nl,D}$ in CrBr$_3$/graphene heterostructures shows an abrupt slowdown as the temperature exceeds the transition temperature $T_C$ (~37 K) of CrBr$_3$, which attributes to the CrBr$_3$'s phase change from ferromagnetic to paramagnetic state.

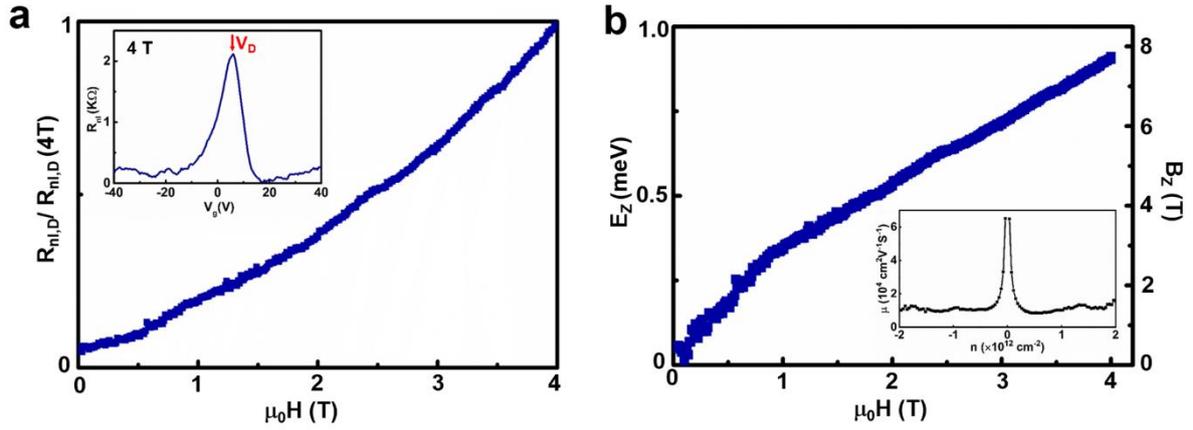

**Figure 3.** Magnetic field dependence measurement and the theoretical calculation of the MPE induced Zeeman splitting energy in CrBr$_3$/graphene van der Waals heterostructures. (a) The plot of normalized $R_{nl,D}$ (using $R_{nl,D}$ measured at 4 T as reference) with external magnetic field sweep. It can be seen that $R_{nl,D}$ rises sharply with external magnetic field increasing, reaches up to orders-of-magnitude enhancement. Inset curve is $R_{nl}$ as a function of $V_g$ under 4 T. (b) Calculated Zeeman splitting energy $E_Z$ and corresponding Zeeman field $B_Z$ is plotted against the external magnetic field, respectively. The inset shows field-effect mobility $\mu_*$ as a function of carrier density of CrBr$_3$/graphene heterostructures.

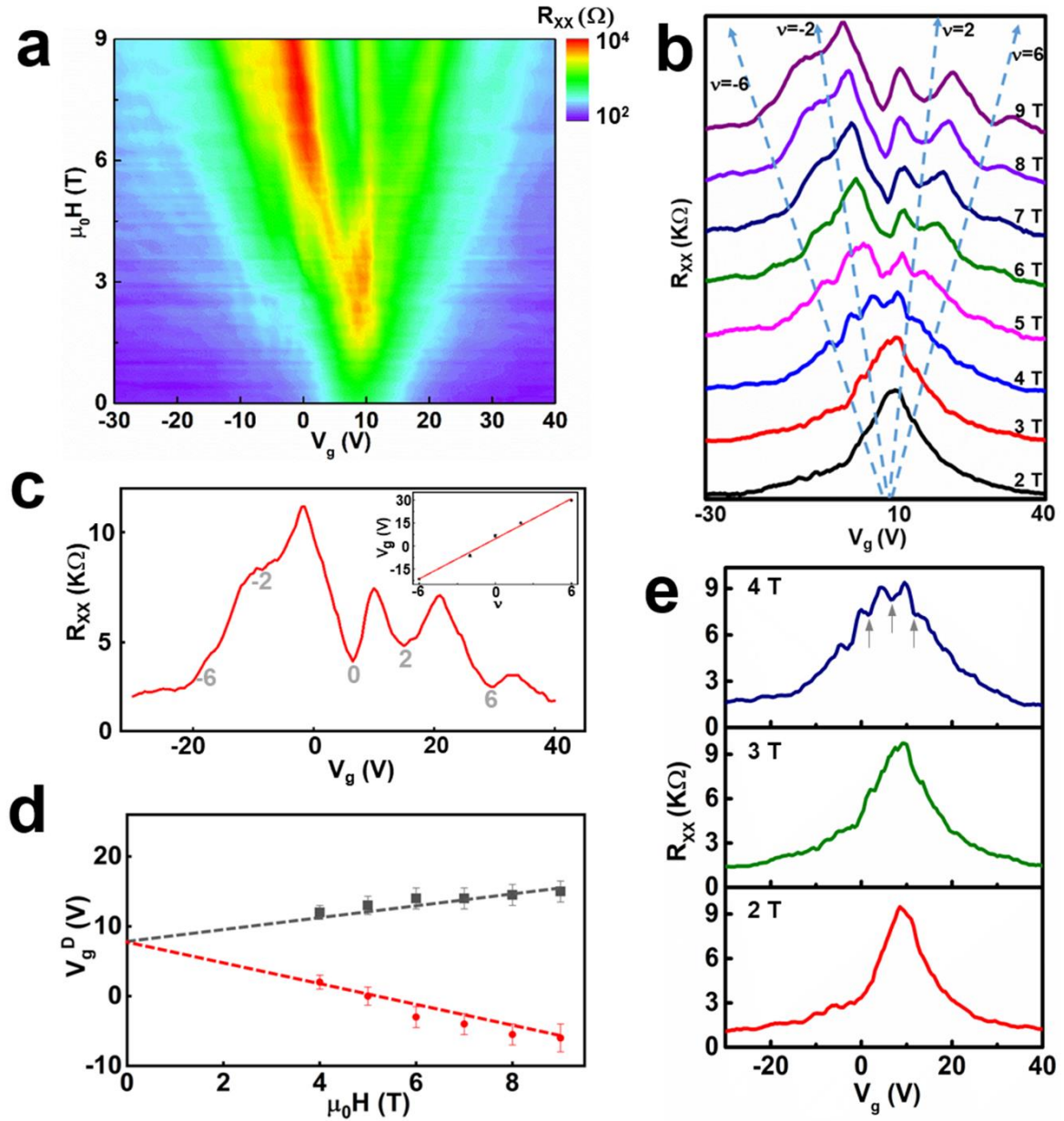

**Figure 4.** MPE effect in the quantum Hall regime of CrBr$_3$/graphene heterostructures. (a) Color map shows Landau Fan of the longitudinal resistance $R_{XX}$ as a function of the gate bias $V_g$ and the applied magnetic field $\mu_0 H$, which is measured at 2 K. (b) Extracted longitudinal resistance under external field from 2 to 9 T from (a). For clarity of labelling the Landau level filling factors, resistance curves are vertically shifted and are proportional to the applied magnetic field. (c) Longitudinal resistance $R_{XX}$ measured at 9 T shows pronounced quantum oscillations. $\nu$ is directly derived from the longitudinal resistance versus gate bias curve at each local $R_{XX}$

minimum, and all $\nu$ are labelled in light grey. Inset figure manifests $V_g$ is linear to each $\nu$. (d) The theoretical predictions (dashed lines) and experimental results (solid scatters) of gate bias to infer the Dirac point position by using capacitance model at $\nu = \pm 2$. (e) The close-up of each local $R_{XX}$ maxima and $R_{XX}$ minima of resistance curves taken at 2 T, 3 T, and 4 T, revealing an anomalous resistance dip and peak-splitting features near $\nu = 0$ that develop with the increasing magnetic field, which suggests the lifting of Landau level degeneracy under large $B_z$.

# Supporting information for "Magnetic Proximity Effect in 2D Ferromagnetic $CrBr_3$/Graphene van der Waals Heterostructures"

Chaolong Tang,[†] Zhaowei Zhang,[†] Shen Lai,[†] Qinghai Tan,[†] Wei-bo Gao*[,†,‡]

[†]Division of Physics and Applied Physics, School of Physical and Mathematical Sciences, Nanyang Technological University, Singapore 637371, Singapore

[‡]The Photonics Institute and Centre for Disruptive Photonic Technologies, Nanyang Technological University, 637371 Singapore, Singapore

## S1. Calculation of Zeeman splitting energy $E_Z$ and total Zeeman field $B_Z$.

In pure graphene, a few publications have reported that Zeeman spin Hall effect (ZSHE) induces spin-up electrons and spin down holes due to graphene's gapless band structure, the Lorenz force further split the electron and hole states and drive this spin current perpendicular to the charge current along the Hall bar contacts[1-4]. Therefore, the ZSHE resulted from spin current can be probed by this non-local charge transport.

Theoretically, the non-local resistance predicted for the ZSHE is[1]

$$R_{nl} \equiv \frac{dV_{nl}}{dI_B} \propto \frac{1}{\rho_{xx}} \left(\frac{\partial \rho_{xy}}{\partial \mu} E_Z\right)^2 \tag{1}$$

where $R_{nl}$ is defined as the non-local resistance, $V_{nl}$ is the non-local voltage drop between the hall bar contacts, $I_B$ is the charge current driven across the Hall bar, μ is the chemical potential, $\rho_{xx}$ and $\rho_{xy}$ is the longitudinal and Hall resistivity, respectively; $E_Z$ is the Zeeman splitting energy.

While the non-local resistance at Dirac point $R_{nl,D}$ can be further written as[1,5]

$$R_{nl,D} = R_0 + \beta(B_0) E_Z^2 \tag{2}$$

where $R_0$ stands for the non-local resistance at zero-field, $\beta$ represents the $\rho_{xx}$ and $\rho_{xy}$ manifested parameters.

To directly derive the expression between $E_Z$ and $R_{nl,D}$ at the different external magnetic field, we introduce the parameter $\alpha$, which is determined as[5]

$$\alpha(B) = \sqrt{\frac{\beta(B)}{\beta(B_0)}} = \frac{E_{Z0}}{E_Z} \cdot \sqrt{\frac{R_{nl,D}(B) - R_0}{R_{nl,D}(B_0) - B}} \tag{3}$$

For evaluating $E_Z$ conveniently, we similarly choose a proper magnetic field $B_0$ as a reference to calculate $E_Z$. While $\beta$ has the following form based on the given formulas above[4],

$$\beta = \frac{\omega}{2l_s} e^{-l/l_s} \frac{1}{\rho_{xx,D}} \left(\frac{\partial \rho_{xy}}{\partial \mu}\right)^2 \bigg|_{\mu=\mu_D} \tag{4}$$

Where $\rho_{xx}$ and $\rho_{xy}$ are magnetic field dependent, other parameters are not. By considering the behavior at charge neutral point, $\rho_{xx}$ is a sum of the Drude-Lorentz resistivity and the electron-hole drag contribution, and thus $\rho_{xx,D}$ is estimated as[2]

$$\rho_{xx,D} = \frac{m_T}{2n_T e^2 \tau}(1 + \tau^2 \Omega^2) \tag{5}$$

$$n_T = \frac{\pi}{12}\frac{K_B^2 T^2}{\hbar^2 v_0^2} \tag{6}$$

Here $m_T \approx 3.29 K_B T/v_0^2$, $n_T$ is the density of thermally activated electrons (holes) at the charge neutral point, $\tau$ is the scattering time, and $\Omega$ is the cyclotron frequency, $v_0$ is Fermi velocity. By substituting $\beta$ back into the parameter $\alpha$, other magnetic field independent parameters are cancelled and this item is left in the form of $\sqrt{\frac{1+(\tau\Omega_{c,B_0})^2}{1+(\tau\Omega_{c,B})^2}}$, which equals to $\sqrt{\frac{1+(\mu_* B_0)^2}{1+(\mu_* B)^2}}$, the mobility $\mu_*$ of the sample is about 10000 cm²/VS in CrBr₃/graphene heterostructures. As to $\rho_{xy}$, $\frac{\partial \rho_{xy}}{\partial \mu} = \frac{\partial \rho_{xy}}{\partial n} \cdot \frac{\partial n}{\partial \mu}$, and $\left.\frac{\partial \rho_{xy}}{\partial n}\right|_{n=0} = \frac{B}{4ecn_T^2}$, $\frac{\partial n}{\partial \mu} = \frac{2\ln 2}{\pi}\frac{K_B T}{\hbar^2 v_0^2}$. Finally, the parameter $\alpha$ can be written as

$$\alpha(B) = \sqrt{\frac{\beta(B)}{\beta(B_0)}} = \sqrt{\frac{\frac{1}{\rho_{xx,D}(B)}\left(\frac{\partial \rho_{xy}(B)}{\partial \mu}\right)^2\Big|_{\mu=\mu_D}}{\frac{1}{\rho_{xx,D}(B_0)}\left(\frac{\partial \rho_{xy}(B_0)}{\partial \mu}\right)^2\Big|_{\mu=\mu_D}}} = \sqrt{\frac{1+(\tau\Omega_{c,B_0})^2}{1+(\tau\Omega_{c,B})^2}} \cdot \sqrt{\frac{\left(\frac{\partial \rho_{xy}(B)}{\partial \mu}\right)^2\Big|_{\mu=\mu_D}}{\left(\frac{\partial \rho_{xy}(B_0)}{\partial \mu}\right)^2\Big|_{\mu=\mu_D}}} = \sqrt{\frac{(1+\mu_*^2 B_0^2)B^2}{(1+\mu_*^2 B^2)B_0^2}} \cdot \tag{7}$$

$$E_Z = \frac{E_{z0}}{\alpha}\frac{\sqrt{R_{nl,D}(B)-R_0}}{\sqrt{R_{nl,D}(B_0)-R_0}} = E_{z0} \cdot \frac{\sqrt{(1+\mu_*^2 B^2)B_0^2}}{\sqrt{(1+\mu_*^2 B_0^2)B^2}} \cdot \frac{\sqrt{R_{nl,D}(B)-R_0}}{\sqrt{R_{nl,D}(B_0)-R_0}} \tag{8}$$

$$B_Z = \frac{E_Z}{g\,\mu_B} \tag{9}$$

Therefore, we could calculate $E_Z$ and $B_Z$ based on the deduced formula above and field dependence of $R_{nl,D}$ data in CrBr₃/graphene heterostructures.

**S2. Observed additional ground states of graphene in CrBr₃/graphene heterostructures.**

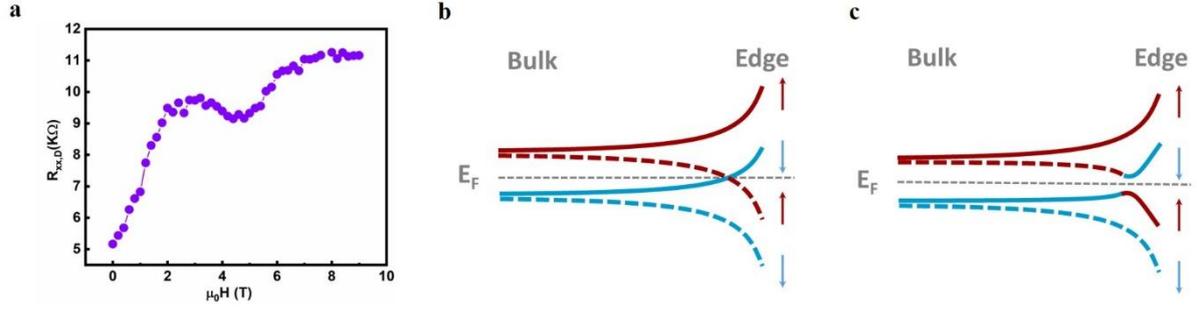

**Figure S1**. Ferromagnetic state to a canted antiferromagnetic state. (a) longitudinal resistance at the Dirac point $R_{XX,D}$ evolves with the applied magnetic field, which initially increases to its peak with increasing magnetic field then drops continuously after 3 T. notably, $R_{XX,D}$ rises gradually again after 5 T. (b) Scheme of the spin-polarized ferromagnetic phase as described in quantum Hall ferromagnetism theory. The sample edge is in gapless state and there exist counter-propagating edge channels with opposite spins, leading to decreasing $R_{XX,D}$. (c) Schematic of the canted antiferromagnetic phase with the opening of an edge state gap.

Another observation associated with the MPE induced $B_Z$ is the change of anomalous longitudinal resistance at the Dirac point $R_{XX,D}$. As shown in Figure S1a, $R_{XX,D}$ initially increases with applied field until its peak at about 3 T and then drops continuously with field increasing, the behavior of which is so different with the reported monotonic increase of $R_{XX,D}$ in graphene/AlOx and other graphene heterostructures[6-8]. While the MPE effect mainly affects the $v = 0$ states, therefore, the intriguing behavior near $v = 0$ is strongly related to the novel electronic structure of graphene ground state. Nevertheless, such unusual properties of CrBr$_3$/graphene heterostructures at $v = 0$ can be suitably described by the spin-polarized ferromagnetic phase in quantum Hall ferromagnetism theory[9-13] as shown in Figure S1b. In detail, the bulk gap at the Dirac cone is dominated by the Zeeman splitting energy $E_Z$ through the sub-Landau level of the spin-up cone and spin-down cone crossover, the sample edge state is gapless and exist counter-propagating edge channels with opposite spins, leading to decreasing $R_{XX,D}$ as agrees with our observation. After applied magnetic field exceeds 5 T, a dramatic increase of $R_{XX,D}$ occurs, indicating additional states forming at the Dirac point with increasing $B_Z$ and suggesting an energy gap is opened at $v = 0$. This feature is also distinct from conventional graphene samples without magnetic coupling. However, the canted antiferromagnetic phase as described in Figure S1c from quantum Hall ferromagnetism theory could appropriately explain the abnormal ground state phase transformation of graphene, stemming from the spin-polarized ferromagnetic phase to canted antiferromagnetic phase transformation occurs with the opening of an edge state gap after 5 T.

Since the ground state phases of the CrBr$_3$/graphene heterostructures is determined by the competition between valley isospin anisotropy energy and Zeeman splitting energy: if Zeeman splitting energy is larger than the doubled valley isospin anisotropy energy, the spin-polarized ferromagnetic phase will present; if less, it will go to canted antiferromagnetic phase[9-13]. While the valley isospin anisotropy energy only relies on the applied magnetic field, Zeeman splitting energy is attributed to the MPE induced field and the applied magnetic field. By tuning the applied magnetic field, we can switch the graphene ground state with different phase between ferromagnetic state and canted antiferromagnetic state at $\nu = 0$.

## S3. Other sample's non-local measurements

In order to verify the accuracy and repeatability of observed physics phenomenon, we fabricated and measured other samples under the same conditions. As shown in Figure S2, similarly, the non-local resistance $R_{nl}$ at Dirac point in CrBr$_3$/graphene heterostructures shows a huge increase as the external magnetic field goes up, confirming a distinct Zeeman splitting energy enhancement in CrBr$_3$/graphene heterostructures.

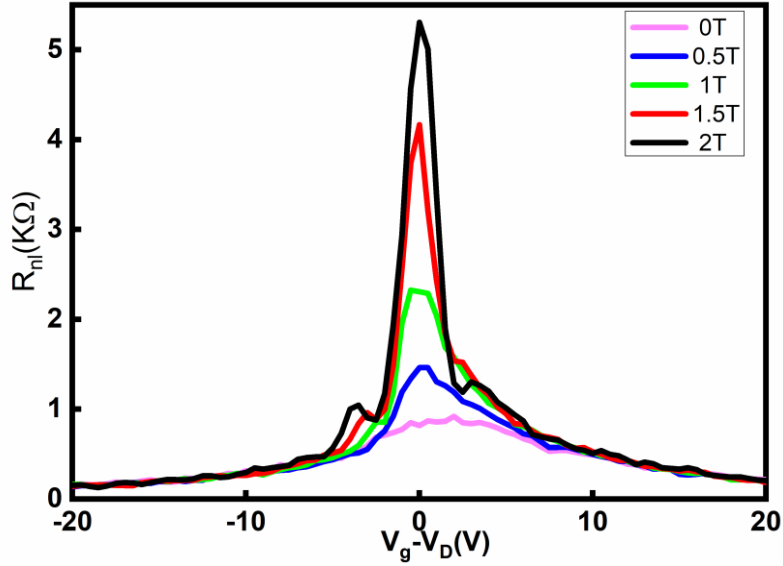

**Figure S2**. Non-local resistance measurement of another device with the same heterostructures and configurations.

### S4. Other sample's four-probe longitudinal resistance measurements

As the magnetic proximity exchange field has a strong influence on the quantum state of graphene, we further examined the heterostructures by measuring the longitudinal resistance in the quantum Hall regime. As shown in Figure S3, longitudinal resistance of graphene $R_{XX}$ in CrBr$_3$/graphene heterostructures shows typical quantum oscillation, which is achieved at a low magnetic field. Insert shows $V_g$ is linear fit to each $v$. Furthermore, the close-up of local $R_{XX}$ minima of resistance curves taken at 2 T and 3 T as shown in Figure S4, reveal extra dip features which develop with the increasing magnetic field, especially near $v = 0$ (marked by red dash circle in the Figure). This peak-splitting feature usually indicates the lifting of Landau level degeneracy under large $B_Z$.

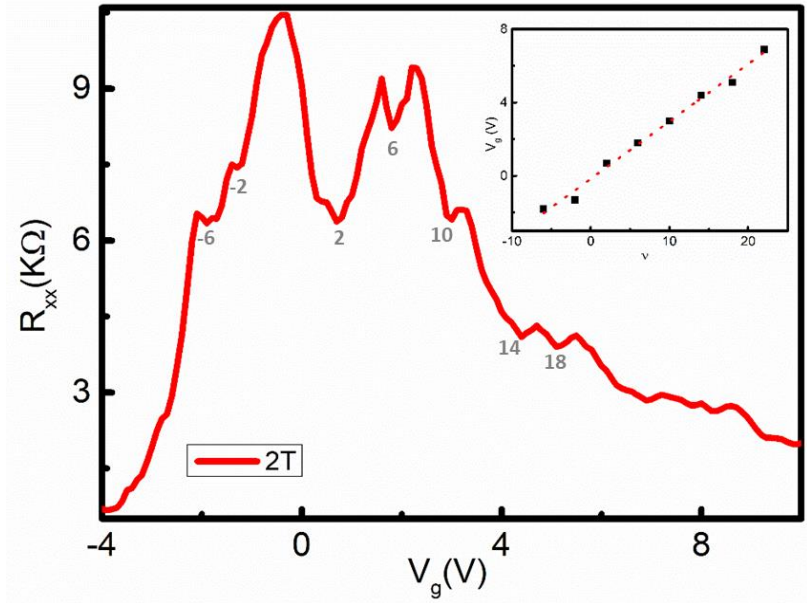

**Figure S3**. Longitudinal resistance $R_{XX}$ measured at 2 T as a function of gate bias, showing typical quantum oscillation in CrBr$_3$/graphene heterostructures. Insert shows $V_g$ is linear fit to each $\nu$.

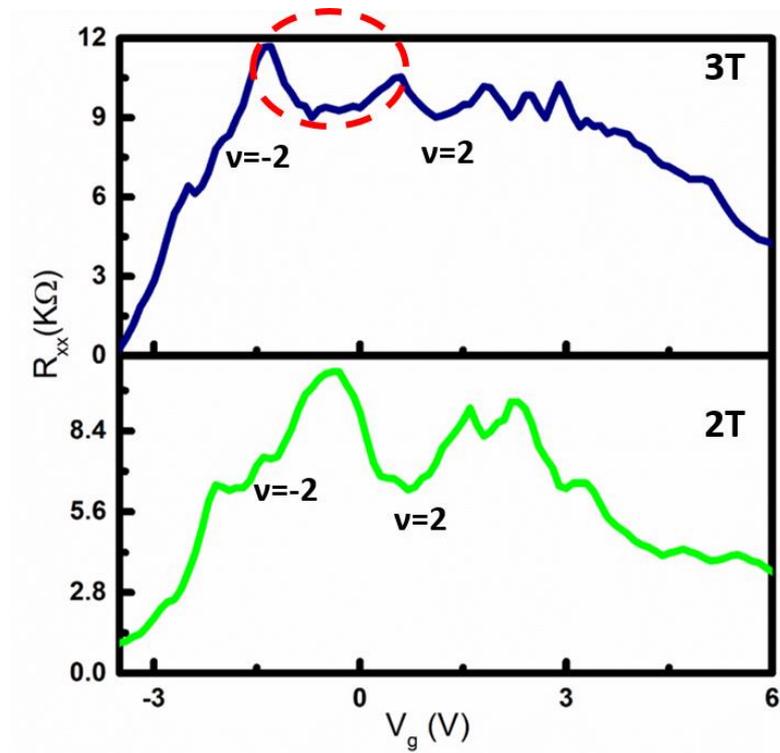

**Figure S4**. The close-up of local $R_{XX}$ minima of resistance curves taken at 2 T and 3 T, revealing extra dip features which develop with the increasing magnetic field, especially near $\nu = 0$ (marked by red dash circle in the Figure).

**S5**. **Possible sources of producing nonlocal resistance $R_{nl}$**

The non-local measurements in the configuration of Figure 2b may involve additional voltage signals except from ZSHE. Thus, we considered possible extrinsic sources that may contribute to $R_{nl}$, they are analyzed and found out to be very small comparing with ZSHE signals so as to affect entire nonlocal resistance less and can be reasonably neglected in this measurement. First, CrBr$_3$ is an insulating ferromagnet, no electrons would prefer to go through it when coupled to a highly conductive graphene in our configuration unless in a magnetic tunnel junction design. The other considerable source to induce voltage signal in nonlocal resistance $R_{nl}$ is the Ohmic contribution which is due to classical diffusive charge transport to the measured nonlocal resistance for a typical Hall bar structure with channel length of $L$ and width of $W$. The Ohmic contribution can be further calculated by van der Pauw formula[14] $R_{nl} = \frac{\rho_{xx}}{\pi} \exp(-\pi \frac{L}{W})$, where the resistivity $\rho_{xx} = R_L \frac{W}{L}$, By using the $R_{XX}$ data measured in Figure 4a and Hall bar geometry $L/W$=3.3 in Figure 1d, we compared the nonlocal resistance resulting from ohmic contribution with the measured nonlocal resistance $R_{nl}$ in Figure S5. Obviously, the measured $R_{nl}$ is four orders of magnitude larger than the calculated Ohmic contribution. We therefore exclude the Ohmic contribution as the origin of the observed $R_{nl}$.

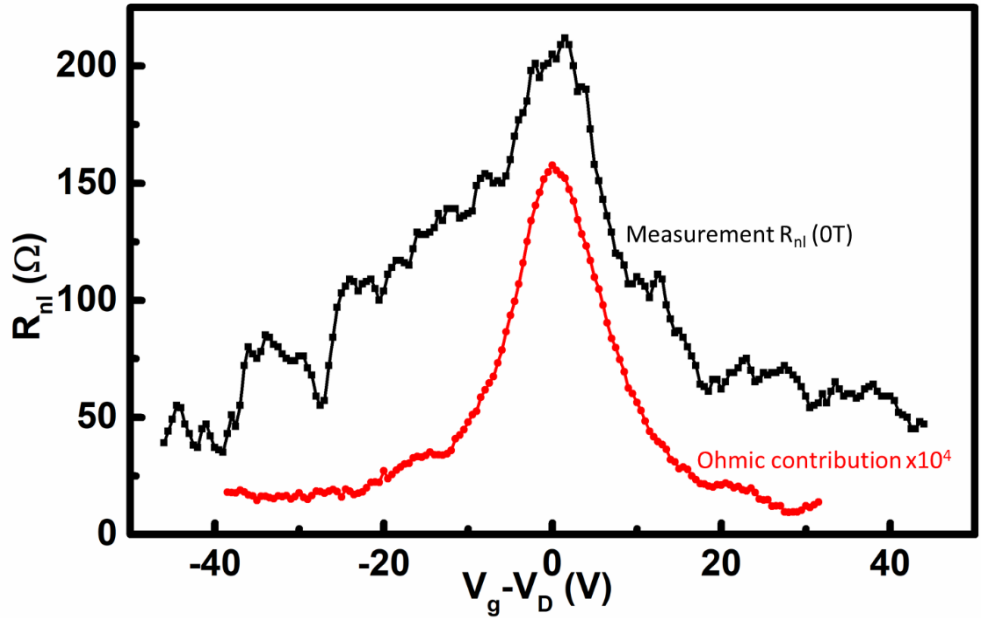

**Figure S5**. Comparison between the measured $R_{nl}$ (black) and a calculation of the Ohmic contribution (magnified $10^4$ times) (red). The Ohmic contribution curve is calculated using the $R_{XX}$ data along Hall bar channel measured at 0 T.